\documentclass{aa}
\usepackage{graphicx}
\begin{document}

\title{Quark-Nova}

\author{R. Ouyed\inst{1}
\and J. Dey\inst{2}
\and M. Dey~\inst{3}
}

\institute{Nordic Institute for Theoretical Physics, Blegdamsvej 17,
DK-2100 Copenhagen, Denmark \and
Department of Physics, Maulana Azad
College, Calcutta 700 013, India \and
Department of Physics, Presidency
College, Calcutta 700 073, India}

\offprints{ouyed@nordita.dk}

\date{Received/Accepted}

\abstract{
We  explore the scenario where 
the core 
of a neutron star (having experienced
a transition to an up and down quark phase)
shrinks into the equilibrated quark object after reaching
strange quark matter saturation density (where 
a composition of up, down and strange quarks is the favored state of matter). 
The overlaying (envelope) material free-falls  following the core 
contraction releasing upto $10^{53}\ {\rm ergs}$ in energy as radiation,
partly as a result of the conversion of envelope material to quarks. This 
phenomena, we named Quark-Nova, leads to a wide
variety of ejectae ranging form the Newtonian, ``dirty"
to the ultra-relativistic fireball. 
The mass range of the corresponding compact remnant
(the quark star) ranges from less than $0.3M_{\odot}$ up to a solar mass. 
We discuss the connection between Quark-Novae and Gamma ray bursts  and
 suggest the recently studied GRB011211 event as a plausible Quark-Nova
 candidate.
}
\maketitle

\section{Introduction}

Recent and fresh observational
data collected by the new generation of X-ray and $\gamma$-ray
satellites suggest that the compact
objects associated with the X-ray pulsars, the X-ray bursters,
particularly the SAX J1808.4-3658 (Li et al. 1999)
are good quark stars candidates. 
If quark stars do exist in nature then how do they form? 
They have been speculated to form directly during
or shortly after some supernova (SN) explosion
when the central density of the proto-neutron star
is high enough to induce phase conversion to quark matter
(e.g., Dai, Peng, \& Lu 1995; Xu, Zhang, \& Qiao 2000). 
Another plausible scenario involves converting neutron stars into
quark stars (Olinto 1987; Cheng \& Dai 1996; Bombaci \& Datta 2000).
Here we explore the scenario where the quark matter  
core contracts and shrinks to the corresponding 
stable compact/quark object.
While this has already been considered in the literature 
(e.g., Rosenhauer et al. 1992; Ma \& Xie 1996),
in our model the core shrinks as to physically separate from the overlaying
 material (we refer to as the envelope).
This approach, also leading to quark star formation,  offers a new and richer
dynamics and allows us to develop the concept of Quark-Nova (QN).

\section{Strange matter mixture and quark stars}

Strange matter, or {\it (u,d,s)}, mixture (a composition of up,
down and strange quarks, hereafter SQM) 
is traditionally modeled with
an equation of state (EOS) based on the MIT-bag model
(Farhi \& Jaffe, 1984). Here, for
illustration purposes, we choose to use the EOS as suggested in
Dey et al. (1998) with  the binding energy 
versus density of such a system shown in Figure 1.  One  finds that from 
about 3$\rho_{\rm N}$ to 7$\rho_{\rm N}$ ($\rho_{\rm N}$ is the
nuclear matter saturation density) the {\it (u,d,s)} mixture has  more binding 
energy than $^{56}$Fe saturating around
$5\rho_{\rm N}\equiv \rho_{\rm ss}$ with an energy about $B_{\rm conv.}\sim 50$ MeV per baryon less than $^{56}$Fe and is  therefore very stable. 

The possible existence of quark stars
(specifically of strange star; SSs), is a direct consequence of the
above described conjecture  that SQM may be the absolute ground
state of strong interacting matter rather than $^{56}$Fe
(Bodmer 1971; Witten 1984; Alcock et al. 1986).
Figure 2 shows the SS Mass-Radius plane resulting from the EOS described in
Dey et al. (1998). SSs can acquire masses up to $1.44M_{\odot}$,
with radii up to 7.06 km (for non-rotating stars). 
However there is no lower limit on their
size (mass) since they would be bound by the
strong interaction even in the absence of gravity.
Among their features 
(Alcock et al. 1986; Glendenning 1997) that are relevant to our present model:

i) The density at the surface  changes abruptly from zero to $\rho_{\rm ss}$.
 The abrupt change (the thickness of the quark surface) occurs within
 about 1 fm, which is a typical strong interaction length scale.

ii) The electrons being bound to the quark matter by the electro-magnetic
 interaction and not by the strong
force, are able to move freely across the quark surface
extending up to $\sim 10^3$ fm above the surface of the star.
Associated with this electron layer is a
strong electric field ($5\times 10^{17}$ V/cm)
which would prevent ionized matter from coming into
direct contact with the surface of the star.
Note that free neutrons (or neutral matter
in general) can easily penetrate the
Coulomb barrier and gravitate to the surface of
the star; they are readily absorbed and converted to quark matter.

\section{Quark-Nova Collapse}

From the calculations by Gentile et al. (1993), the phase transition 
to a state with the up and down quarks would take place in the inner core
(with mass $M_{\rm c} < 0.3M_{\odot}$ and radius $R_{\rm c} < 1$ km)
where matter is homogeneous ($\sim \rho_{\rm c}$); it is this portion
of the star which we discuss next.

\subsection{Core contraction}

Take a neutron star (hereafter NS, of mass $M_{\rm NS}$)  which
experienced an increase in their central density
above deconfinement values
($\rho_{\rm d}$; that is $\rho_{\rm c} > \rho_{\rm d}$)
--  following a spin-down
history or as a direct result of a SN
explosion (see Ma 2002 for a recent discussion). 
Conversion of {\it (u,d)} 
matter to {\it (u,d,s)} takes place via weak interactions  where  
non-leptonic (for example, $u+d \rightarrow u+s$) process is of greater 
rate (Anand et al., 1997; Olinto 1987, Heiselberg et al. 1991). 
If the density of the core does not exceed
$\rho_{\rm ss}$ then it will never
undergo  transition into the lower energy branch of the matter.
 The NS remains a contaminated stable neutron star namely, a hybrid star 
with a core density lying between $\rho_{\rm d}$ and $\rho_{\rm ss}$ (e.g.
Chapt. 8 in Glendenning 1997). Only when $\rho_{\rm ss}$ is
reached can a QN occur.
We mention that while in the case of a proto-NS the
trapped neutrinos will shift the possible onset of the phase transition
 to SQM to higher densities, with respect to the case of a cold and
 neutrino-free NS, the basic idea presented below remains the same.

The speed of propagation of the conversion/contamination front and its
 dynamics are limited by the low weak rate of
at which the quarks diffuse. As such, the front propagation might
be considered slow compared to the conversion 
(the conversion into SQM occurs within a very short period
of time $\sim 10^{-7}$ s; Dai et al. 1995) leading to the
interesting situation where the core density exceeds  $\rho_{\rm ss}$ 
much before the rest (upper layers) of the star. A
scenario where the core  would shrink to the corresponding stable
bare SQM object.

\begin{figure}[t!]
\centerline{\includegraphics[width=0.4\textwidth]{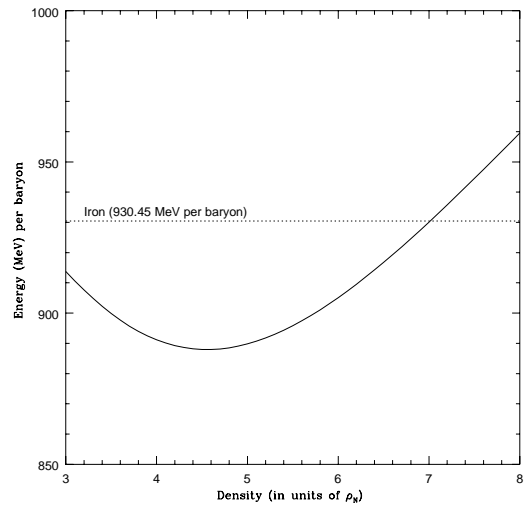}}
\caption{Energy per baryon for SQM 
 in terms of the density  as modeled
by Dey et al. (1998).  The horizontal dotted line shows the
energy per baryon in $^{56}$Fe. 
At high density ($\sim 5\rho_{\rm N}$), SQM 
with its lower energy is the preferred state of matter.
}
\label{Fig.1}
\end{figure}

The radius of the contracted core we can estimate
as $R_{\rm ss}/R_{\rm core}\simeq (\rho_{\rm c}/\rho_{\rm ss})^{1/3}$.
Given our fiducial values
$\rho_{\rm ss}\simeq 5\rho_{\rm N}$ and $\rho_{\rm c}\sim 2\rho_{\rm N}$
(for a canonical $1.44M_{\odot}$ and standard 10 km radius generic
neutron star) the core of the NS would shrink by  as much as 30\%. 
The maximum gravitational energy released is thus 
$(0.3M_{\odot}c^2)\Delta R/R_{\rm c} \sim 10^{53}$ ergs; although 
very large\footnote{ Detailed calculations which includes the various details
of the neutron star and SS 
structural properties can be found in Bombaci \& Datta (2000;
their Figure 2 in particular) leading to similar numbers.} would be mostly
carried out by neutrinos and only a very tiny part of it
would be released as radiation (Janka \& Ruffert 1996). 

\subsection{Envelope collapse and the neutron drip sphere}

Because of the density contrast between the core 
and the overlaying material ($\sim \rho_{\rm N}$), the envelope
free-fall time ($t_{\rm ff,env.}$) is longer than the core
contraction time ($t_{\rm ff,ss}$). 
Simple considerations imply $t_{\rm ff,env.}/t_{\rm ff,ss}
\sim \sqrt{\rho_{\rm ss}/\rho_{\rm N}}\sim \sqrt{5}$
and the core physically separates from the overlaying material.
The infalling envelope material would consist
of the neutral part ($\rho_{\rm env.} > \rho_{\rm drip}=4.3\times 10^{11}$
g/cc with $\rho_{\rm drip}$ the neutron drip density)
the ionized plasma part 
of the envelope material 
($\rho_{\rm env.} < \rho_{\rm drip}$) and the upper solid layers
(the crust).
The Coulomb barrier of the newly formed
quark star which is of the order of $E_{\rm C}\simeq 15$ MeV
(Alcock et al. 1986) makes the QN collapse very intricate. 
Indeed, the charged envelope material (including the crust)
will be subject to the tremendous Coulomb polarizing force.
One might argue that the kinetic
energy of a proton when accreted ($\sim Gm_{\rm p}M_{\rm ss}/2R_{\rm ss}\sim 100$ MeV where $m_{\rm p}$ is the proton mass and $G$
the gravitational constant) is large enough
to overcome such a barrier, however, it is likely that the kinetic
energy will be radiated away via heat (Frank et al. 1992) before
reaching the neutron drip sphere.
In this case the energy per accreted proton
is reduced below the Coulomb barrier as to halt
the infalling material leaving it suspended (``polarized up") 
a few hundred Fermi above the surface of the SQM core.
As we show below, the energy released during conversion of the
neutrons is enough to expel any ionized material 
that might have been left
suspended above the newly formed SS as to leave it naked/bare.

\begin{figure}[t!]
\centerline{\includegraphics[width=0.4\textwidth]{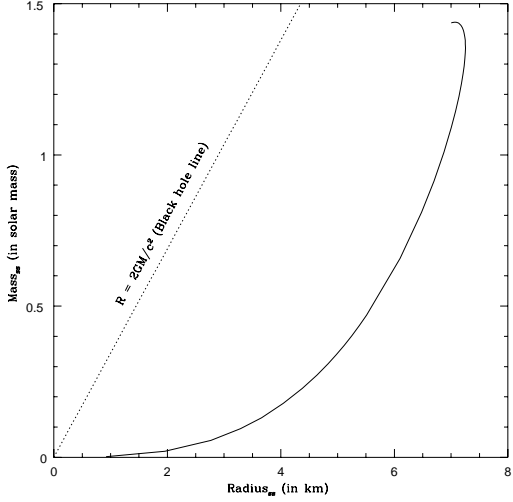}}
\caption{The Mass-Radius relation for
non-rotating SSs (Dey et al. 1998).
}
\label{Fig.2}
\end{figure}

\section{Energetics}

During envelope collapse we shall have tremendous energy release as radiation.
First, is the gravitational binding energy
which is released during the first stage of the collapse
(onto the neutron drip line), followed by the conversion energy. 
The neutrons not subject to the Coulomb barrier 
will continue their fall and be be converted to SQM; 
{\it the envelope is essentially squeezed out of its neutral component}.

\subsubsection{Accretion energy}

The accretion energy can be written as,
\begin{equation}
E_{\rm acc.} \simeq \eta_{\rm acc.} M_{\odot} c^2~({M_{\rm env.}\over M_{\odot}})
({R_{\rm Schw.,ss}\over R_{\rm ss}})\ .
\end{equation}
In general, $R_{\rm Schw.,ss}/ R_{\rm ss}\simeq 1/2$ (Figure 2) which leave
us with,
\begin{equation}
E_{\rm acc.} \simeq {\eta_{\rm acc.} \over 2}M_{\odot}c^2~({M_{\rm env.}\over M_{\odot}})\ .
\end{equation}

Since the drip line radius does not refer to a `hard' surface but only
to a region into which matter can fall before it is squeezed out of
its neutrons, it is far from clear on the value of $\eta_{\rm acc.}$ one needs
to adopt. It might seem appropriate as a first approximation to take
the accretion efficiency of SSs to lie close to that of
black holes ($\eta_{\rm BH}\sim 0.1$) and  of neutron stars ($\eta_{\rm NS} \sim
0.15$); that is $\eta_{\rm acc.}\sim 0.1$.

\subsubsection{Conversion energy}

Each neutron converted to SQM is accompanied by a
photon of energy $B_{\rm conv.}$.
The conversion energy is therefore 
\begin{equation} 
E_{\rm con.} \simeq  {M_{\rm env.}\over (m_{\rm n}-B_{\rm nuc.}/c^2)}\ B_{\rm conv.}\ {\rm MeV}\ , 
\end{equation} 
where $m_{\rm n}$ is 
the baryon mass and $B_{\rm nuc.}$ 
the nuclear binding energy. Equation above is best written as 
\begin{equation} 
E_{\rm con.} \simeq  \eta_{\rm conv.} \left ( {M_{\rm env.}\over M_{\odot}} 
\right ) 
M_{\odot}c^2, 
\end{equation} 
where $\eta_{\rm conv.}= B_{\rm conv.}/(m_{\rm n}c^2-B_{\rm nuc.})$
 is the {\it strangeness conversion efficiency} 
 estimated to be $\eta_{\rm conv.}\sim 0.1$ given 
$B_{\rm conv.}\simeq 50$ MeV.

The energy released in the QN is thus
\begin{equation}
E_{\rm QN} = \eta_{\rm QN}
M_{\odot}c^2~\left ({M_{\rm env.}\over M_{\odot}}\right )\ ,
\end{equation}
where 
\begin{equation}
\eta_{\rm QN} =\left ({\eta_{\rm acc.}\over 2}
 + \eta_{\rm conv.}\right )\ .
\end{equation}
Thus,
\begin{equation} 
E_{\rm QN} \simeq
2\times 10^{53}\ {\rm ergs}
\left ( {M_{\rm env.}\over M_{\odot}} 
\right ) 
\left ( {\eta_{\rm QN}\over 0.15}\right )\ ,
\end{equation}
which can be as high
as $\simeq 10^{53}$ ergs.

\section{Quark-Nova features}

\subsection{QN ejecta and compact remnant}

The fireball's radiation energy density $a T^4$ ($T = B_{\rm conv.}
\sim 50$ MeV) by far exceeds that of the gravitational energy density
in the envelope ($G M_{\rm ss}\rho_{\rm env.}/R_{\rm ss}$).
Some envelope material and any matter that might ended  suspended
above the neutron drip sphere is thus loaded into the fireball
making up the QN ejecta.
The corresponding Lorentz factor can be written as   
\begin{equation}
\Gamma_{\rm QN} ={E_{\rm QN}\over M_{\rm ejec.}c^2}\sim ({\eta_{\rm QN} \over \nu})\ ,
\end{equation}
where $\nu$ defines the amount of ejected material ($M_{\rm ejec.} = \nu M_{\rm env.}$).
The QN ejecta could lead 
to a wide variety of ejectae (with different Lorentz factors) ranging from a
 Newtonian, ``dirty" to an ultra-relativistic fireball.

The compact remnant (the quark star) mass is mainly dependent on 
the amount of envelope material that falls
onto the core and acquires a mass in the $M_{\rm c}<M_{\rm ss}<M_{\rm NS}$
 range.
An efficient envelope extraction would
lead to a light star with $M_{\rm ss} < 0.3M_{\odot}$.
Note that any fallback material from the QN ejecta
would end up as a fossil disk material\footnote{The newly formed quark star
would have spun up during the contraction
naturally offering some angular momentum.}
which if  channeled to the polar caps could later form a crust
(recall that the disk material is globally charged; Xu et al. 2000).

\subsection{QN rate}
 
The QN rate can be expressed as
\begin{equation}
R_{\rm QN} = \zeta\ R_{\rm Ma}
\end{equation}
where 
\begin{equation}
 R_{\rm Ma} \simeq 10^{-5}\ \left( \frac{P_{\rm i}}{20\ \rm{ms}}\right)
\left( \frac{R_{\rm NS}}{10^{-2}} \right) \ \rm{yr}^{-1}\ \rm{galaxy}^{-1}\ ,
\label{qnrate}
\end{equation}
is the rate at which NSs undergo a phase transition
to quark matter in the core (see Ma \cite{ma02}).
$R_{\rm NS}$ is the neutron star birth rate and $P_{\rm i}$
the initial spin of a neutron star.
The factor $\zeta$ represents the fraction of 
neutron stars that will undergo
the transition to the $\rho_{\rm ss}$ density following deconfinement
in the core. We expect $\zeta$ to be small making the QN a very rare
event. A value for $\zeta$ is suggested below.

\subsection{The Gamma Ray Burst connection?}

The QN as we have seen would lead
to the formation of hot quark stars (the collapse and conversion
process would heat up the star to high temperatures).
In Ouyed \& Sannino (2002) and Ouyed (2002, for a recent review) 
we explained how Gamma ray bursts (GRBs) 
might result from the evaporation of such stars into photons.
Such a process as we showed is triggered when
quark matter phase transitions come into play at the surface
of the star.
As the quark star (with surface temperatures above
a few tens of MeV) cools it enters the
so-called 2 flavor color superconductivity (2SC) phase
where glueballs decay into photons (the fireball;
Ouyed \& Sannino, 2001) consuming most of the star in this process.
It is interesting to note from Eq.~\ref{qnrate} that $\zeta \sim 0.1$
would imply QN rate of {\it one per million year per galaxy}
in agreement with the GRB rate derived from 
BATSE\footnote{The Burst and Transient Source Experiment
 detector on the COMPTON-GRO satellite observes on 
average one burst per day. This corresponds,
with the simplest model -- assuming no cosmic evolution of the rate -- to
about once per million years in a galaxy (Piran 1999).}
measurements. This might be an indication that this
aspect of our model warrants further study.

\subsection{The QN-SN connection: {\bf GRB 011211}?}

If deconfinement (and later $\rho_{\rm ss}$) happens late following a 
spin-down history then the collapse would occur much after the SN event
 with no QN-SN interaction. On the other hand,
if deconfinement happens during or shortly after the SN event
(e.g. for very massive progenitors),
the SN is then expected to be followed by the QN.  
The time delay between the two events (the time it takes the core
to reach the critical density $\rho_{\rm ss}$) could
vary from seconds to days depending on the conditions
in the core of the proto-NS before contraction (such as the temperature and 
density) and the details of the spread
of contamination/conversion front.

Eventually, the QN-ejecta catches up with the dense SN-ejecta which is 
heated up producing emission. 
QNe would thus be detected indirectly by their effect on the preceding
 SNe ejectae. In the case where the QN-ejecta
consists of a hot fireball ($\nu << 1$) and the 
QN occurs within a few days after the SN, the latter
is caught up when the expanding QN-ejecta is in its X-ray phase
(the fireball afterglow phase; Piran 1999).
The dense SN-ejecta when heated up would then produce X-ray emission. 
Such a signature might have been observed in the 
GRB011211 case (Reeves et al. 2002) where the authors
conclude that the measured X-ray emission is best explained as
an emission from SN material illuminated by a GRB which occurred a few days
following the SN explosion. Given its features,
we are tempted to suggest GRB011211 as a plausible QN candidate.

\begin{acknowledgements}
We are grateful to an anonymous referee for the remarks that helped improve
this work.
\end{acknowledgements}

\end{document}